\documentclass[final,3p,times]{elsarticle}

\usepackage{wrapfig}
\usepackage{epsfig}
\usepackage{amsthm}
\usepackage{hyperref}
\usepackage{amsmath}
\usepackage{cleveref}
\usepackage{CJKutf8}
\usepackage{latexsym}
\usepackage{amstext}
\usepackage{amsfonts}
\usepackage{dsfont}
\usepackage{color}
\usepackage{amssymb}
\usepackage{relsize}
\usepackage{amsxtra}
\usepackage{verbatim}
\usepackage{hyperref}
\usepackage{graphics}
\usepackage{graphicx}
\usepackage{multirow}
\usepackage{ifpdf}
\usepackage{lineno}

\usepackage{caption}
\usepackage[labelfont=bf]{caption}
\usepackage[labelsep=space]{caption}
\usepackage[nodots, compress]{numcompress}
\biboptions{sort&compress}

\usepackage{xcolor}
\usepackage{hyperref}
\hypersetup{
colorlinks,
citecolor=[violet],
linkcolor=[red],
urlcolor=[blue]
}

\journal{Journal of Energy Storage}

\begin{document}

\begin{frontmatter}

\title{Predicting doping strategies for ternary nickel-cobalt-manganese cathode materials to enhance battery performance using graph neural networks}

\author[1]{Zirui Zhao}
\author[2]{Dong Luo}
\author[3,4]{Shuxing Wu}
\author[1]{Kaitong Sun}
\author[3,4]{Zhan Lin\corref{cor1}}
\ead{zhanlin@gdut.edu.cn}
\author[1]{Hai-Feng Li\corref{cor1}}
\cortext[cor1]{Corresponding author}
\ead{haifengli@um.edu.mo}
\affiliation[1]{organization={Institute of Applied Physics and Materials Engineering, University of Macau},
            addressline={Avenida da Universidade, Taipa},
            city={Macao SAR},
            postcode={999078},
            country={P.R. China}}
\affiliation[2]{organization={Hunan Provincial Key Laboratory of Advanced Materials for New Energy Storage and Conversion, School of Materials Science and Engineering, Hunan University of Science and Technology},
            city={Xiangtan},
            postcode={411201},
            country={P.R. China}}        
            
\affiliation[3]{organization={Jieyang Branch of Chemistry and Chemical Engineering Guangdong Laboratory},
            city={Jieyang},
            postcode={515200},
            country={P.R. China}}
                
\affiliation[4]{organization={School of Chemical Engineering and Light Industry, Guangdong University of Technology},
            city={Guangzhou},
            postcode={510006},
            country={P.R. China}}

\begin{abstract}
The exceptional electrochemical performance of lithium-ion batteries has spurred considerable interest in advanced battery technologies, particularly those utilizing ternary nickel-cobalt-manganese (NCM) cathode materials, which are renowned for their robust electrochemical performance and structural stability. Building upon this research, investigators have explored doping additional elements into NCM cathode materials to further enhance their electrochemical performance and structural integrity. However, the multitude of doping strategies available for NCM battery systems presents a challenge in determining the most effective approach. In this study, we elucidate the potential of ternary NCM systems as cathode materials for lithium-ion batteries. We compile a comprehensive database of lithium-ion batteries employing NCM systems from various sources of prior research and develop a corresponding data-driven model utilizing graph neural networks to predict optimal doping strategies. Our aim is to provide insights into the NCM-based battery systems for both fundamental understanding and practical applications.
\end{abstract}

\begin{graphicalabstract}
\includegraphics[width=0.88\textwidth]{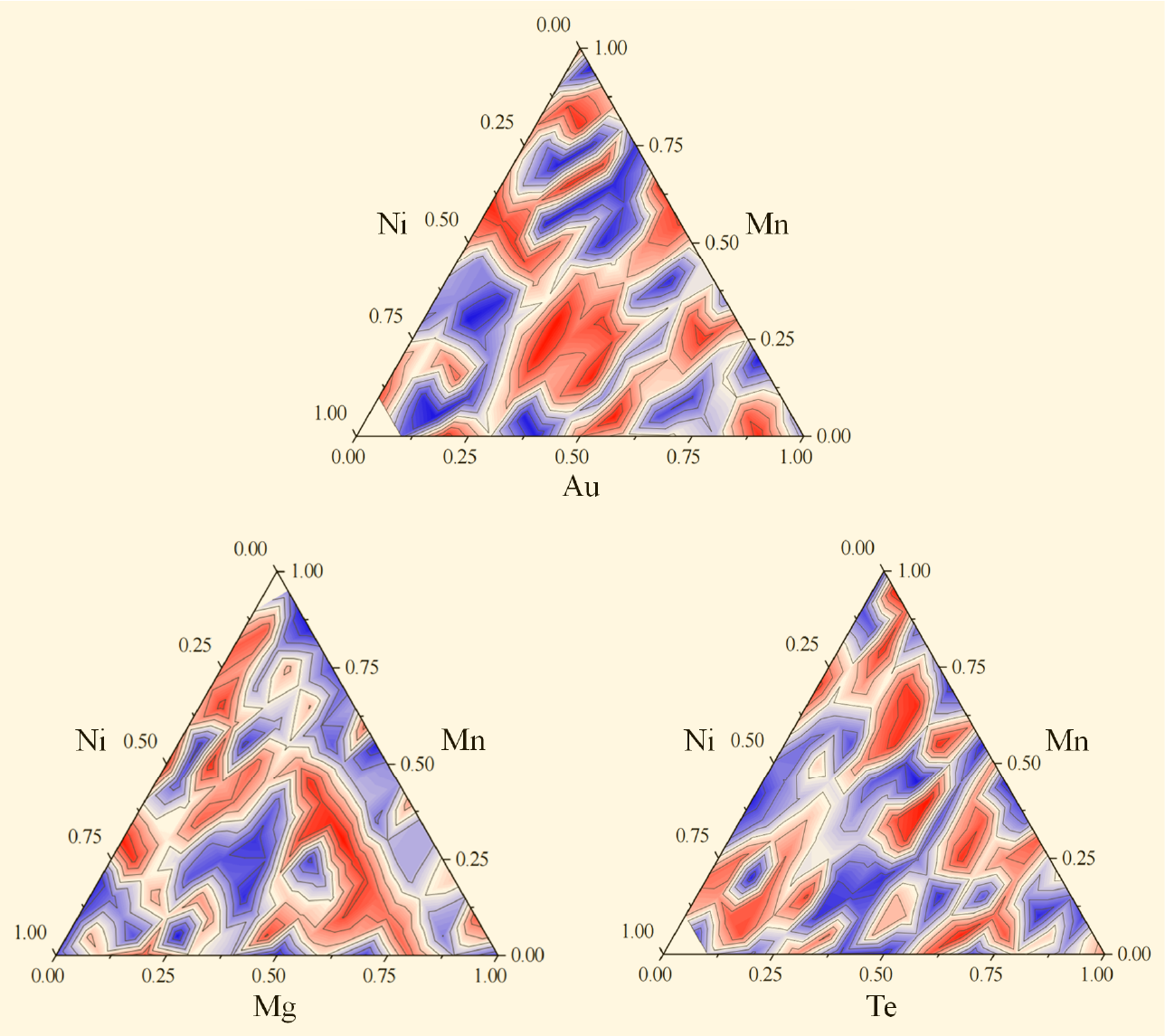} \\
\medskip
\noindent{\textbf{Caption of Graphical Abstract:} Phase diagrams depict doping ratios' impact on performance enhancement of $[\text{Li}]_{3a}[\text{NiMn}]_{3b}[\text{O}_{2}]_{6c}$ cathode materials in Li-ion batteries, featuring gold (Au), magnesium (Mg), and tellurium (Te) dopants. Graph neural networks optimize doping strategies for enhanced battery performance.}
\bigskip
\label{GA}
\end{graphicalabstract}



\end{frontmatter}

\section{Introduction}

The ternary nickel-cobalt-manganese (NCM) system, typically comprising different ratios of nickel (Ni), cobalt (Co), and manganese (Mn) ions, has attracted considerable attention as a promising cathode material for lithium-ion batteries due to its favorable electrochemical performance. This performance is characterized by high capacity, stability, and excellent cycle life \cite{goodenough2010challenges}. The crystal structure of NCM battery materials is typically layered, akin to other lithium-ion battery cathodes, with a transition metal (Ni, Co, Mn) oxide layer and a lithium-ion layer. The arrangement of metal ions in the oxide layer plays a crucial role in determining the electrochemical properties of the material \cite{ma2014phosphorus}.

Despite their promising characteristics, NCM batteries face challenges that hinder their widespread commercialization. Notably, their energy density remains lower compared to batteries using alternative cathode materials \cite{ren2010influence, zhou2003lithium}. Additionally, rapid performance degradation during cycling poses a significant concern, attributed to structural instability, electrolyte decomposition, and side reactions with electrode materials \cite{ji2012sns, ran2015porous}. Safety concerns, particularly regarding thermal runaway and fire risk, further highlight the challenges associated with NCM-based batteries \cite{li2016nitrogen}. To address these challenges, researchers are exploring various strategies, including surface coatings, doping elements, and new electrolyte formulations, to enhance the performance and durability of NCM-based batteries \cite{slater2013sodium, armand2008building, han2015free}. 

Among these strategies, doping additional elements into NCM cathode materials is particularly effective in addressing the inherent challenges of lower energy density and enhancing overall performance. At the atomic level, doping modifies the electrochemical properties and structural stability of NCM materials, yielding several key benefits. Firstly, doping can increase the energy density of NCM cathodes by introducing elements that facilitate higher lithium-ion mobility and storage capacity. For instance, certain dopants create additional active sites for lithium-ion intercalation. Secondly, doping improves cycling stability by stabilizing the crystal structure during charge and discharge cycles. Elements such as aluminum or magnesium are known to reduce cation mixing and prevent undesirable phase transitions, thereby enhancing the material’s longevity and performance consistency. Thirdly, doping can enhance the thermal stability of NCM cathodes by reducing the material’s propensity for exothermic reactions at high temperatures, thereby mitigating the risk of thermal runaway. Moreover, doping helps reduce the degradation of NCM materials by mitigating the effects of high-voltage cycling and preventing the dissolution of transition metals into the electrolyte, thus prolonging the battery’s operational lifespan \cite{ZHOU2023101248, choi2012challenges, landi2009carbon}. These findings demonstrate that doping is a viable and impactful method for enhancing the performance of NCM cathode materials in next-generation lithium-ion batteries.

However, current doping strategies rely heavily on empirical evaluation and theoretical calculations, which may not fully capture the complexities of practical operating conditions \cite{sevilla2014energy, yu2024high, zhu2016composite, kesavan2020design, sun2013carbon}. To overcome these limitations, we propose leveraging data-driven deep learning models. Graph neural networks (GNNs) have emerged as powerful tools for predicting material properties based on microstructural properties \cite{cha2021nanoporous, liu2021bismuth, bruna2013spectral}. Utilizing their ability to characterize molecular structures through high-dimensional matrices, artificial GNN models accurately capture intricate relationships between atoms and predict material properties.

In this study, we aim to evaluate the electrochemical performance of NCM batteries containing dopants using a data-driven GNN model. By training the model on a dataset of NCM structures with different dopants, we seek to develop a predictive model capable of accurately assessing battery quality. This research not only advances the application of deep learning techniques in materials science but also provides an efficient method for evaluating cathode quality in battery systems.

\section{Methods}

\subsection{Graph neural networks}

GNNs are a category of artificial neural network models designed to work on graph-structured data. They have been widely applied in materials science due to their ability to characterize relationships between elements in high-dimensional matrix data and respond effectively to their overall effect \cite{bruna2013spectral}. Moreover, GNNS consistently offer several significant advantages over traditional machine learning models, particularly in materials science, where input data often consists of two-dimensional matrices representing various properties and interactions. 

Firstly, GNNs are adept at capturing intricate relationships and dependencies between nodes in a graph, which is crucial for accurately modeling systems where interactions between entities, such as atoms in a material, are as important as the entities themselves. Secondly, GNNs can efficiently handle large and complex graphs by leveraging local connectivity patterns and sparsity, making them scalable to large datasets typical in materials science. Furthermore, for many tasks involving graph data, GNNs have demonstrated superior performance compared to traditional models, particularly in predicting material properties and behaviors. Additionally, GNNs offer flexibility in handling various types of graph data, including directed, undirected, weighted, and unweighted graphs. These advantages make GNNs a powerful tool for analyzing and learning from graph-structured data in materials science, leading to improved accuracy and insights.

For example, we consider \( G = (V, E) \) to indicate a graph, where \( V \) is the set of nodes representing entities, and \( E \) is the set of edges indicating relationships between nodes. Each node \( v_i \in V \) is associated with a feature vector \( x_i \), and each edge \( e_{ij} \in E \) is associated with a weight or label \( w_{ij} \) \cite{gori2005new}.

The basic operating principle of GNNs is to collect information from neighboring nodes and iteratively update the node embedding \cite{chen2014methodology}. This is achieved through a message passing scheme:

\[
h_i^{(l+1)} = \sigma \left( \sum_{j \in N(i)} f \left( h_i^{(l)}, h_j^{(l)}, e_{ij} \right) \right),
\]
where \( h_i^{(l)} \) is the embedding of node \( v_i \) at layer \( l \), \( N(i) \) is the set of neighboring nodes of node \( v_i \), \( f \) is a structured function that combines information from the nodes and edges, and \( \sigma \) is a non-linear activation function. GNNs also contain pooling operations, allowing them to aggregate information from multiple nodes, as expressed in the following equation:

\[
h_{\text{pool}} = \text{POOL} \left( \{ h_i^{(L)} \}_{i \in V} \right),
\]
where \( h_{\text{pool}} \) is the entire graph's pooled expression, and \( \{ h_i^{(L)} \}_{i \in V} \) are the final layer node embedding, and \( \text{POOL} \) is the pooling function.

In our GNN model, we utilized the Rectified Linear Unit (ReLU) as the activation function. The ReLU function, defined as \( f(x) = \max(0, x) \), introduces non-linearity into the model while maintaining computational efficiency. The choice of ReLU is driven by its ability to address the vanishing gradient issue, which may arise with other activation functions like the sigmoid or hyperbolic tangent functions. By facilitating more effective gradient flow during backpropagation, ReLU enhances the training performance and convergence speed of deep learning models, making it particularly suitable for our GNNs to evaluate the impacts of various doping elements on NCM cathode materials.

GNNs excel in mathematically expressing the connections among points in high-dimensional data, enabling the models to learn complex, high-dimensional structured data and generalize to unseen datasets. They efficiently handle image-structured data and extend the predictive power to large high-dimensional structured datasets \cite{chung1997spectral-1}. This characteristic provides a robust framework for learning high-dimensional data with predominantly two-dimensional graphical structures, applicable in fields such as molecular structure prediction and recommender systems.

In contemporary research, GNNs are increasingly utilized in materials science to predict material properties based on atomic structure. These artificial neural network models represent molecules as a two-dimensional matrix, with atoms as nodes and bonds as edges, effectively describing the microscopic components and structural properties of materials. GNNs can learn intricate patterns in molecular structures and accurately predict properties like energy, stability, and reactivity, which are highly correlated with the materials' structural properties.

GNNs play a crucial role in predicting new materials with similar structural properties based on known material crystal structures. They aid in designing improved catalysts by modelling the atomic structure and interactions with reactant molecules to predict catalytic activity and selectivity. In the domain of battery material prediction and design, GNNs predict capacity, cycle stability, and rate capability by learning the atomic structure of electrode materials, crystallographic information, and interactions with lithium ions. This approach streamlines battery material research, providing valuable insights and accelerating discoveries in the field. 

Wang \emph{et al}. proposed a graph convolutional neural network (GCNN) based on semiconductor crystal structure to predict bandgaps. The GCNN model achieves high accuracy in predicting the bandgap of various materials, which is superior to traditional machine learning models \cite{wang2023capacity}. This approach not only accelerates the discovery of new semiconductors with desirable bandgaps but also provides insight into the fundamental structure-property relationships of materials \cite{verma2018graph}.

\subsection{Data collection and processing}

Several key steps in the data collection and processing for predicting material properties using GNNs are outlined below. First, a dataset is constructed containing the atomic structure of the material and its corresponding properties. The electrochemical properties are proportionally converted and aggregated to a maximum score of 100 points, establishing a linkage with the structural information for calibration. Each material is represented as a graph \( G = (V, E) \), where \( V \) is the set of nodes representing atoms, and \( E \) is the set of edges representing bonds. The graph is constructed based on the atomic structure, with each node \( v_i \) associated with a feature vector \( x_i \) representing atomic properties, and each edge \( e_{ij} \) associated with a feature vector \( w_{ij} \) representing bond properties \cite{zhou2003lithium, ren2010influence, sevilla2014energy, landi2009carbon, choi2012challenges, goodenough2010challenges, ma2014phosphorus, ji2012sns, li2016nitrogen, han2015free,yu2024high}. 

One the dataset is constructed, it is divided into three parts: the training set, validation set, and test set, each organized in separate tables. The GNN model employs a gradient descent algorithm to propagate predictions from the dataset to each node, optimizing the training update parameters. Training is performed on the training set to learn the mapping from the high-dimensional data representation to the material properties \cite{chen2014methodology}. Models are evaluated on validation and test sets using metrics such as mean square error (MSE). Ultimately, the trained model predicts specific electrochemical properties of the material, providing insights into the best doping strategy and offering valuable guidance for material design and discovery.

\begin{figure*} [!t]
\centering \includegraphics[width=0.82\textwidth]{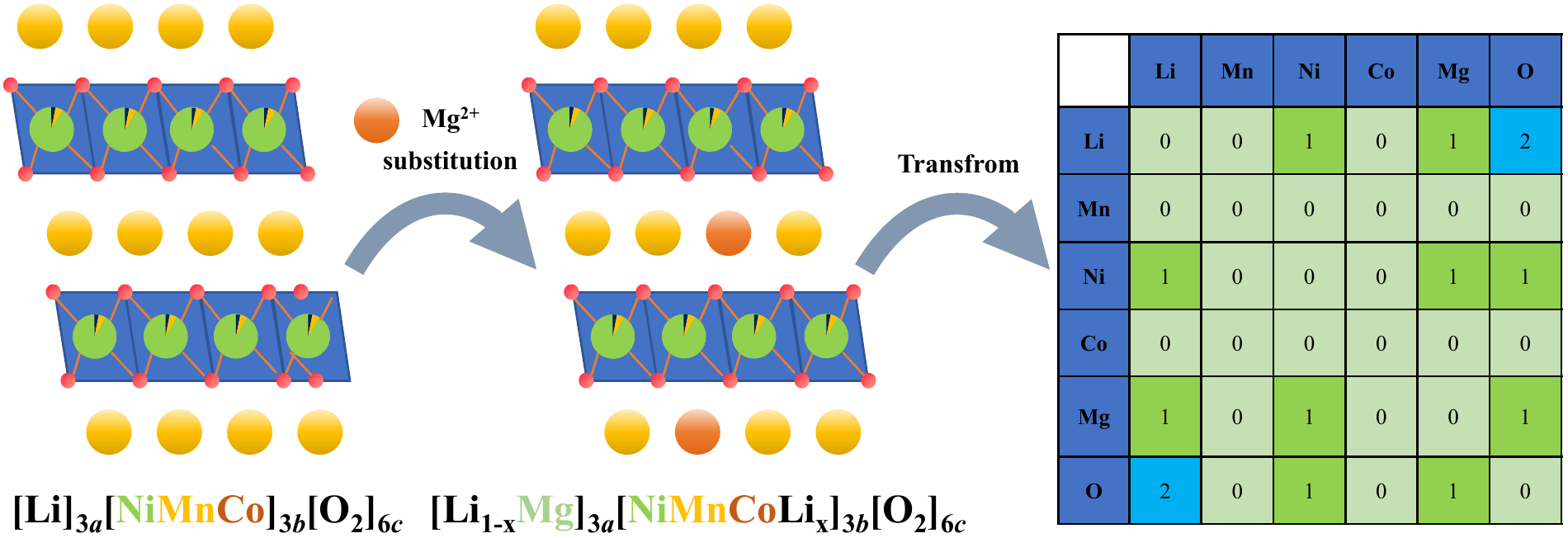}
\caption{Illustration depicting the transformation process for converting crystallographic data of the cathode materials into a matrix format suitable for recognition by an artificial graph convolutional neural network model.}
\label{Matrix}
\end{figure*}

To better recognize and process crystallographic data, the transformation of crystallographic data for recognition by an artificial GCNN can be illustrated using the example of $[\text{Li}]_{3a}[\text{NiMnCo}]_{3b}[\text{O}_{2}]_{6c}$, as depicted in Fig.~\ref{Matrix}. In this representation, each element is assigned a numerical value and arranged in a matrix. The connections (chemical bonds) between elements are denoted by integers, facilitating subsequent computational operations within the GCNN model.

\section{Results and discussion}

\subsection{Construction of graph neural network models}

\begin{figure*} [!t]
\centering \includegraphics[width=0.82\textwidth]{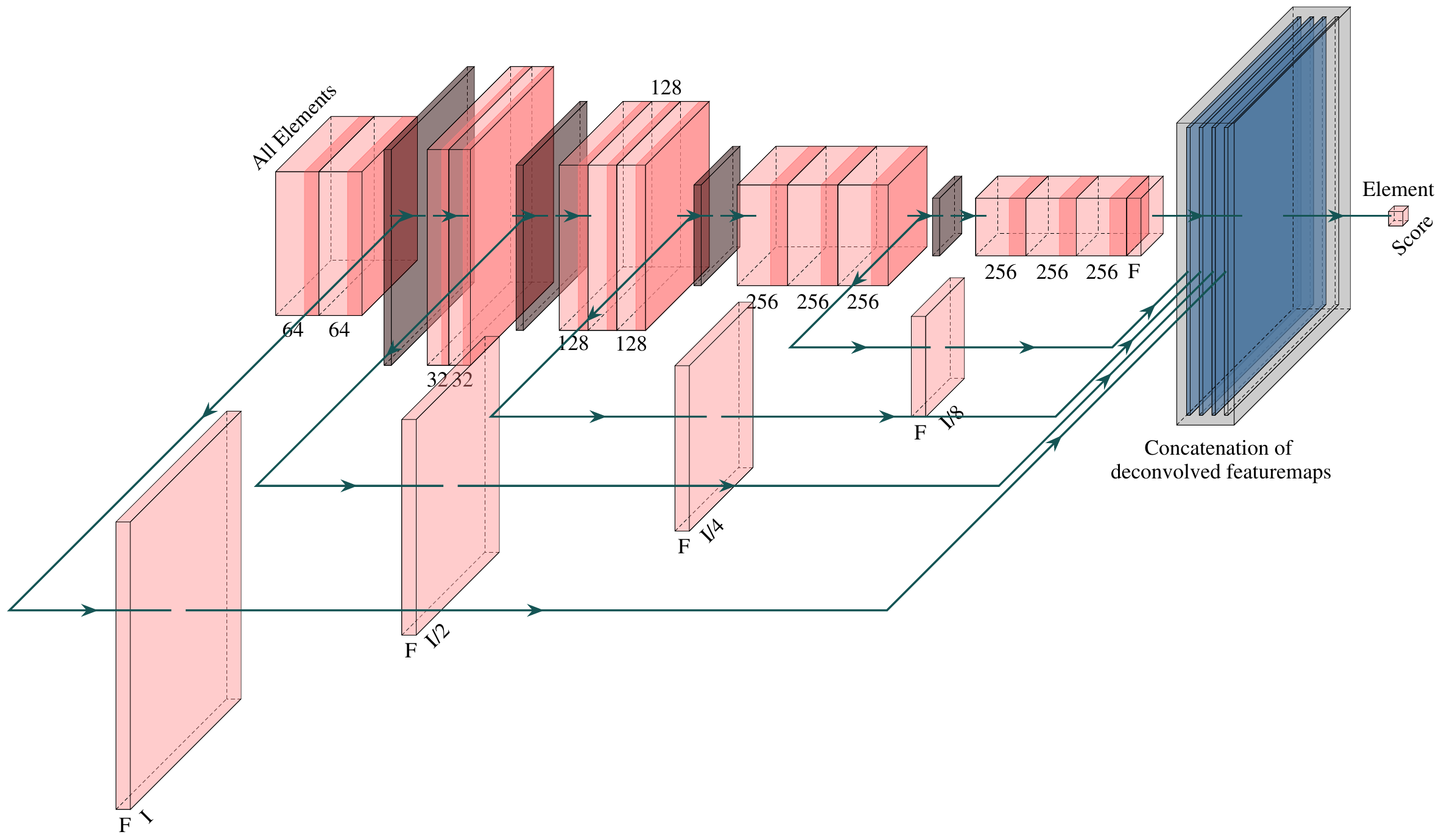}
\caption{Schematic diagram of the artificial graph convolutional neural network architecture, delineating 5 convolutional layers, 4 pooling layers, and 6 fully connected layers.}
\label{application-2}
\end{figure*}

Several key components are involved in constructing a GNN model for predicting material properties. The core of the GNN lines in the graph convolutional layer, which aggregates information from neighboring nodes in the graph. The update rule for the graph convolutional layer is defined as follows:

\[ h_i^{(l+1)} = \sigma \left( \sum_{j \in N(i)} \frac{1}{c_{ij}} W^{(l)} h_j^{(l)} \right), \]
where \( h_i^{(l)} \) represents node \( i \)'s representation at layer \( l \), \( N(i) \) is the set of neighboring nodes of node \( i \), \( c_{ij} \) is a normalization constant for the edge between nodes \( i \) and \( j \), \( W^{(l)} \) is the weight matrix at layer \( l \), and \( \sigma \) is a non-linear activation function. After multiple graph convolutional layers, node representations are aggregated into graph-level representations using readout functions. Common readout functions include sum, mean, or maximum pooling, expressed as:

\[ h_{\text{graph}} = \text{READOUT} \left( \{ h_i^{(L)} \}_{i \in V} \right), \]
where \( h_{\text{graph}} \) is the graph-level representation, \( \{ h_i^{(L)} \}_{i \in V} \) are the final layer node representations, and \( \text{READOUT} \) is the readout function. To predict materials properties, the graph-level representation is processed through multiple fully connected layers, as shown below:

\[ \hat{y} = \text{FC} \left( h_{\text{graph}} \right), \]
where \( \hat{y} \) is the predicted material property, and \( \text{FC} \) is a fully connected layer. The predicted material property \( \hat{y} \) is compared with the truth property \( y \) using a suitable loss function, such as MSE, as shown below:

\[ \mathcal{L} = \text{Loss} \left( \hat{y}, y \right). \]
Through backpropagation and gradient descent algorithms, GNN models are continuously trained to minimize the loss function. Fig.~\ref{application-2} shows a schematic of our constructed artificial GNNs, comprising 5 convolutional layers, 4 pooling layers, and 6 fully connected layers.

\subsection{Model training and validation}

When training and validating GNN models for predicting material properties, we aim to minimize the empirical risk \( \mathcal{R}_{\text{emp}} \) over the training dataset \( D_{\text{train}} \). This is achieved by updating the parameters \( \theta \) of the GNN model using an iterative optimization algorithm called stochastic gradient descent:

\[ \theta^{(t+1)} = \theta^{(t)} - \eta \nabla_{\theta} \mathcal{L}(\theta), \]
where \( \eta \) is the learning rate, and \( \mathcal{L}(\theta) \) is the loss function. We then use a validation set \( D_{\text{val}} \) to fine-tune hyperparameters and prevent overfitting. After each training epoch, the model's performance is evaluated on the validation set, and hyperparameters are adjusted based on the validation performance to minimize the validation loss \( \mathcal{L}_{\text{val}} \). Additionally, we employ \( k \)-fold cross-validation, with \emph{k} set to 5 in this study. The dataset is divided into \emph{k} folds, and the model is trained and validated \( k \) times using different folds for validation and training. 

To further prevent overfitting, regularisation techniques, such as L1 or L2 regularisation, are applied by adding a penalty term to the loss function:

\[ \mathcal{L}_{\text{reg}} = \mathcal{L} + \lambda \sum_{i=1}^{N} {|} \theta_i {|}^2, \]
where \( \lambda \) is the regularization parameter and \( \theta_i \) are the model parameters. This helps mitigate the internal covariate shift problem and enhances overall training speed and stability.

To evaluate the performance of GNN models for predicting material properties, we use various metrics and comparative analyses. Evaluation metrics such as MSE, mean absolute error (MAE), coefficient of determination (\( R^2 \) score), and accuracy are utilized to quantify the performance of GNN models. These metrics are defined as follows:

\[ \text{MSE} = \frac{1}{N} \sum_{i=1}^{N} (\hat{y}_i - y_i)^2, \]

\[ \text{MAE} = \frac{1}{N} \sum_{i=1}^{N} | \hat{y}_i - y_i |, \]

and \[ R^2 = 1 - \frac{\sum_{i=1}^{N} (\hat{y}_i - y_i)^2}{\sum_{i=1}^{N} (y_i - \bar{y})^2}. \]

By employing these training and validation methods, we effectively train and evaluate GNN models for predicting material properties, ensuring robust and reliable predictions. We focus on five main electrochemical performance metrics during model training: cycle life, sustainability, thermal stability, initial specific energy, and initial weight capacity. These metrics are normalized and combined into a score out of 100 for reference.

\subsection{Prediction results of the performance of cathode materials with various elemental dopants}

In this study, evaluating the quality of cathode materials in the NCM system is crucial for optimising the electrochemical performance of lithium-ion batteries. We assessed the electrochemical performance from various aspects, including the conductivity of the cathode materials. Higher conductivity enhances battery charging and discharging efficiency and power output. Additionally, electrochemical stability is vital to prevent unwanted side reactions and ensure long cycling performance. We considered parameters such as stable charging and discharging periods, as well as the material's ability to withstand high voltages and temperatures without decomposing. Further, compatibility with electrodes is essential to prevent degradation and maintain battery performance. We collected data on cathode-electrode interactions, using 0 and 1 to indicate reaction and non-reaction, respectively, to form the dataset. Doping other elements into cathode materials can significantly alter their characteristics and enhance battery performance. We compiled data from previous studies to train our GNN models, predicting the effects of dopant elements on cathode materials' stability and compatibility with electrodes, and assigning corresponding scores. 

In our approach, the data for the GNNs is processed and formatted to ensure it is both two-dimensional and feature-specific. Initially, we collect raw data on the doping elements and their effects on the NCM cathode materials. This raw data is then preprocessed to construct feature vectors that encapsulate relevant properties of the doping elements, such as atomic size, electronegativity, and valence state. Each feature vector represents a node in the graph, and edges between nodes denote the relationships or interactions between different elements. This structure allows the GNN to effectively learn and predict the properties of the doped materials \cite{Zhoumachine}.
Additionally, we utilize an adjacency matrix to represent the atomic connections within the material. This matrix indicates whether an atomic bond exists between two atoms and specifies the type of bond, with different chemical bonds being encoded by distinct numerical values. This approach ensures that the model accurately captures the complexities of the material's structure and composition.

After training, our model demonstrated commendable performance, achieving an accuracy rate of 93.1${\%}$ and a relatively low MSE of 0.005 on the training data. Furthermore, on the test set, the model exhibited an even higher accuracy rate of 88.3${\%}$ and a notably lower MSE of 0.02. These results indicate that our model is not only precise and reliable during training but also maintains excellent predictive performance on unseen data. The high accuracy and low MSE on both training and test sets underscore the robustness of our approach and its potential applicability in further research on material properties.

\begin{figure*} [!t]
\centering \includegraphics[width=0.82\textwidth]{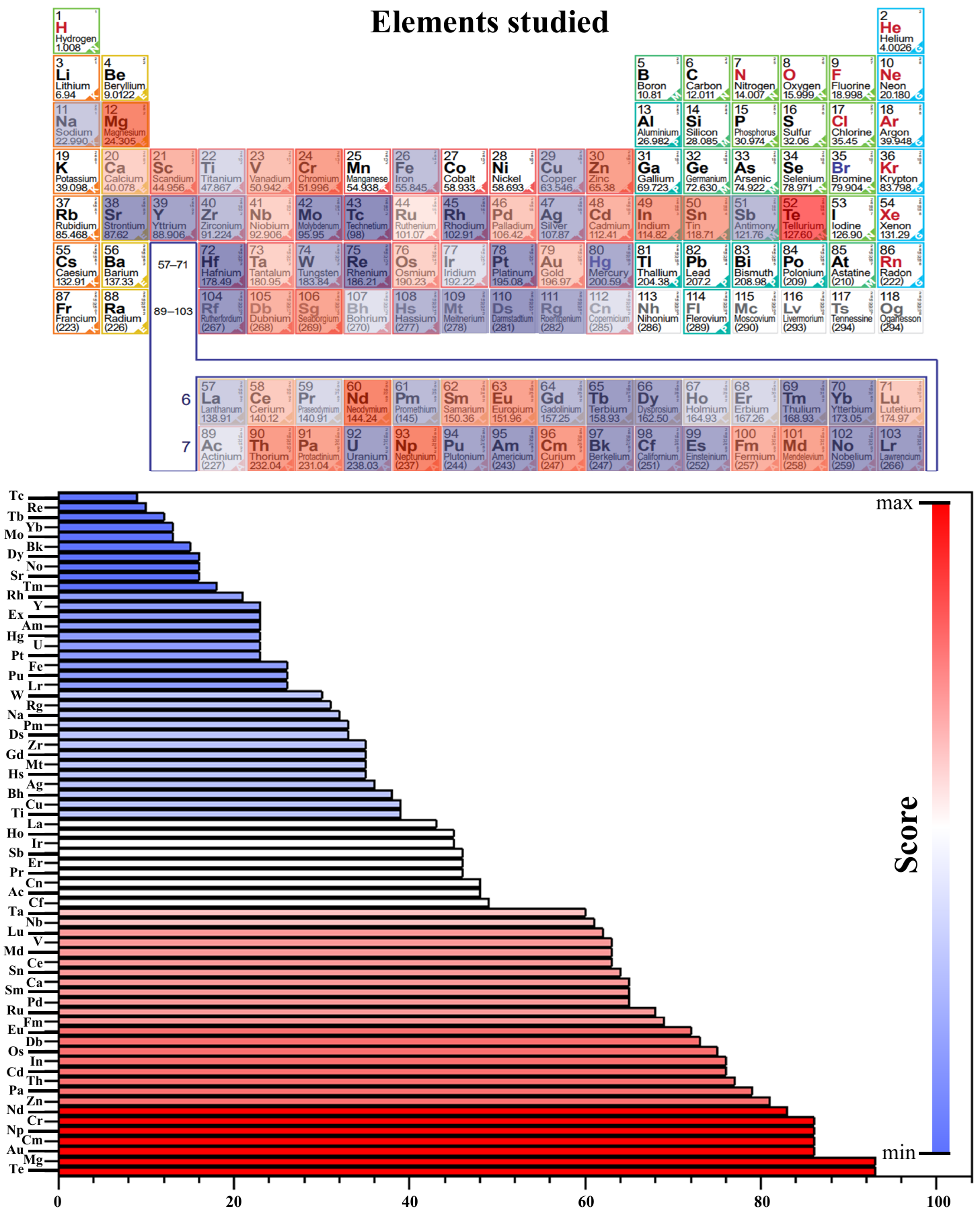}
\caption{Comparison of performance among 73 different elements doped into the $[\text{Li}]_{3a}[\text{NiMn}]_{3b}[\text{O}_{2}]_{6c}$ cathode materials, with emphasis on gold (Au), magnesium (Mg), and tellurium (Te) as the top-performing elements selected for subsequent predictive modeling. Schematic depiction of the periodic table (top panel) sourced from ptable.com, highlighting the masked elements which serve as the doping elements in this study.}
\label{application-3}
\end{figure*}

In recent years, due to cobalt's high cost and toxicity, researchers have shifted focus to cobalt (Co)-free NCM battery systems. Consequently, our calculations exclude cobalt proportions for computational simplicity.
We initially screened 73 different elements for doping into the $[\text{Li}]_{3a}[\text{NiMn}]_{3b}[\text{O}_{2}]_{6c}$ cathode materials. The initial selection of the 73 doping elements was conducted through a systematic and rigorous process. We began by conducting a comprehensive review of existing literature and theoretical studies to identify potential dopants \cite{luo2023li}. Key criteria for selection included atomic size, valence states, and electronic structure, as these properties significantly influence the electrochemical performance and stability of NCM cathode materials. Additionally, elements that have demonstrated promising results in similar material systems or possess unique properties that could potentially enhance lithium-ion battery performance were prioritized \cite{LiuRegula}. This systematic approach allowed us to identify the most promising doping candidates for further detailed study. By adhering to these criteria, we aimed to ensure a robust and scientifically sound basis for our doping element selection, thereby enhancing the reliability and credibility of our research findings. With the model's prediction, gold (Au), magnesium (Mg), and tellurium (Te) showing promising performance enhancements. These elements were then selected for further predictive modeling to assess their impact on cathode properties using the formulation $[\text{Li}_{1-\text{x}}\text{M}]_{3a}[\text{NiMnLi}_{\text{x}}]_{3b}[\text{O}_{2}]_{6c}$. The specific results are summarized in Fig.~\ref{application-3}.

\begin{figure*} [!t]
\centering \includegraphics[width=0.82\textwidth]{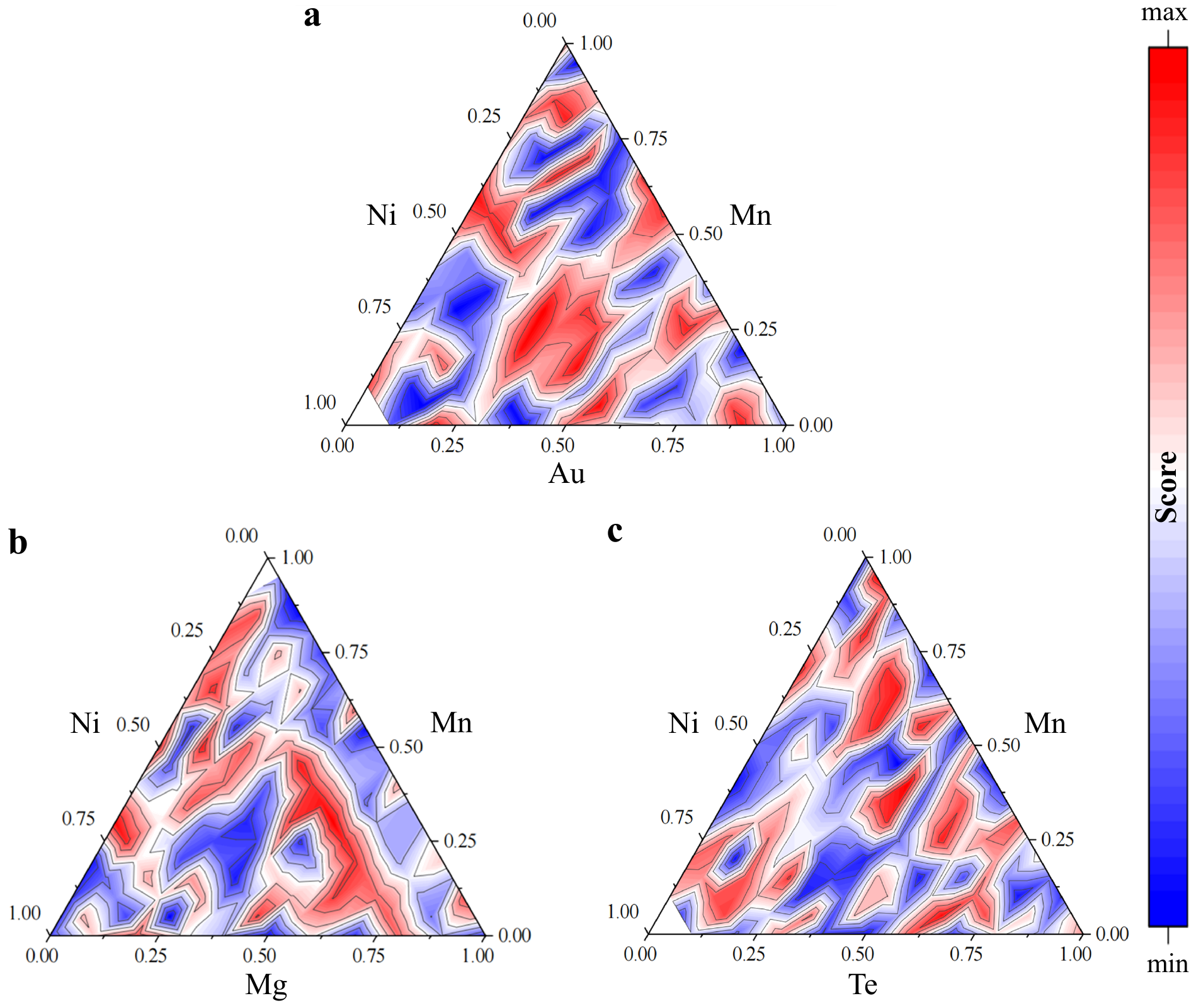}
\caption{Predicted electrochemical properties of doped $[\text{Li}]_{3a}[\text{NiMn}]_{3b}[\text{O}_{2}]_{6c}$ compounds as cathode materials in Li-ion batteries. The doped elements include gold (Au) (a), magnesium (Mg) (b), and tellurium (Te) (c), elucidating the impact of doping ratios on performance enhancement and optimal doping levels.}
\label{application-4}
\end{figure*}

In Fig.~\ref{application-4}, certain regions identified as blank areas were indicative of exceptionally high Ni content. During the model pre-training phase, it was observed that when the Ni content exceeded 90$\%$, and the material was doped with Au and Te, a significant deterioration in cycling performance occurred. To ensure the robustness and validity of our analytical results, we made the decision to exclude combinations with Ni content surpassing 90$\%$. This exclusion was implemented to prevent the inclusion of misleading or unrepresentative data in the compositional analysis, thereby upholding the integrity and accuracy of our research findings.

Moreover, we extracted optimal doping ratios for the best-performing elements among the three and summarized them in Table~\ref{Table-1}. 

GNNs excel in handling high-dimensional matrix datasets, facilitating better predictions of cathode composition data's effects on properties such as conductivity, stability, and compatibility. Hence, we utilized GNNs to comprehensively assess cathode quality in the Ni-Co-Mn ternary system, crucial for predicting overall material performance.

Overall, this study trains a corresponding GNN model and provides a predictive assessment of cathode material quality in NCM batteries based on previous research data. Considering the effects of doping elements on conductivity, stability, and electrode compatibility, we concluded that Au, Mg, and Te are the most suitable doping elements, along with their specific optimized ratios. These findings offer valuable insights for understanding cathode properties and optimizing battery performance.

\section{Conclusions}

In this study, we investigate how GNNs predict the electrochemical performance of cathode materials in NCM battery systems, with a particular focus on cobalt-free cathode materials. We analyze the potential of various doping elements to enhance battery performance. Initially, we provide an overview of the fundamental principles of GNNs, highlighting their capacity to effectively model complex atomic structures in materials. Given the complexity of modeling atomic structures, GNNs offer a promising approach to understanding and predicting the properties of materials at the atomic level.

We then address the current research landscape and challenges in cathode materials for NCM batteries, emphasizing the need for more accurate and efficient methods to guide elemental doping strategies. Accurate modeling is crucial in this context, as it directly impacts the development of high-performance battery materials. Our approach involves evaluating and training GNN models to assess the electrochemical performance of cathode materials using metrics such as MSE, MAE, coefficient of determination, and accuracy. Through this evaluation, we demonstrate the effectiveness of GNNs in predicting material properties. 

Furthermore, we develop an artificial convolutional GNN model tailored specifically for predicting the electrochemical properties of cathode materials in NCM battery systems, considering the influence of various doping elements. By evaluating 73 different doping elements, we identify Au, Mg, and Te as the most promising dopants and determine their optimal doping ratios. These findings have significant implications for optimizing NCM battery performance, particularly in enhancing electrochemical properties. Additionally, they offer practical guidance for designing high-performance cathode materials for next-generation lithium-ion batteries. By bridging the fundamental principles of GNNs with the practical challenges in NCM battery research, our study provides a comprehensive framework for developing advanced battery materials.

\clearpage

\section*{CRediT authorship contribution statement}

\textbf{Zirui Zhao:} Conceptualization, data curation, formal analysis, investigation, methodology, visualization, writing-original draft. \textbf{Dong Luo:} Conceptualization, data curation, formal analysis, investigation, methodology, visualization. \textbf{Shuxing Wu:} Formal analysis, investigation, methodology, visualization. \textbf{Kaitong Sun:} Formal analysis, investigation, methodology, visualization. \textbf{Zhan Lin:} Conceptualization, funding acquisition, methodology, project administration, supervision, visualization, writing-review \& editing. \textbf{Hai-Feng Li:} Conceptualization, funding acquisition, methodology, project administration, supervision, visualization, writing-review \& editing.

\section*{Declaration of competing interest}

The authors declare that they have no known competing financial interests or personal relationships that could have appeared to influence the work reported in this paper.

\section*{Data availability}

Data will be made available on request.

\section*{Acknowledgments}

This work was partially supported by the Guangdong S\&T programme: 2022A0505020027 and 2023A0505020009 and the National Natural Science Foundation of China: 52371217 and 52102221. The work at University of Macau was supported by the Science and Technology Development Fund, Macao SAR (File Nos. 0090{/}2021{/}A2 and 0049{/}2021{/}AGJ) and the Guangdong{-}Hong Kong{-}Macao Joint Laboratory for Neutron Scattering Science and Technology (Grant No. 2019B121205003).

\clearpage

\bibliographystyle{elsarticle-num-names}
\bibliography{ML-3}

\clearpage

\clearpage

\clearpage

\clearpage

\clearpage

\begin{table}[!t]
\caption{Optimized doping ratios of Mg, Te, and Au in $[\text{Li}]_{3a}[\text{NiMn}]_{3b}[\text{O}_{2}]_{6c}$ compounds as cathode materials for Li-ion batteries.}
\label{Table-1}
\setlength{\tabcolsep}{2.8mm}{}
\renewcommand{\arraystretch}{1.0}
\begin{tabular} {ccccc}
\hline
\hline
\text{Doped Element} & \text{Self} & \text{Ni} & \text{Mn} & \text{Total score} \\
\hline
\text{Mg} & 0.2 & 0.5 & 0.3 & 93 \\
\text{Te} & 0.2 & 0.4 & 0.4 & 96 \\
\text{Au} & 0.4 & 0.3 & 0.3 & 98 \\
\hline
\hline
\end{tabular}
\end{table}

\end{document}